\documentclass[tikz,aps,pra,onecolumn]{revtex4}
\usepackage[usenames,dvipsnames]{color}
\usepackage{tikz}
\usepackage{graphicx}
\usepackage{dcolumn}
\usepackage{bm}
\usepackage{amsmath}
\usepackage{amsfonts}
\usepackage{psfrag}
\usepackage[bookmarks,colorlinks=false]{hyperref}
\usepackage{mathtools}
\usepackage{accents}

\usepackage[mathcal]{euscript}

\newcommand*{\I}{i}

\def\bra#1{\mathinner{\langle{#1}|}} 
\def\ket#1{\mathinner{|{#1}\rangle}}

\newcommand{\Eq}[1]{Eq.~(\ref{#1})}
\newcommand{\Eqs}[2]{Eqs.~(\ref{#1})-(\ref{#2})}

\begin{document}

\title{Virtual photons in the ground state of a dissipative system}

\author{Simone \surname{De Liberato}}

\affiliation{School of Physics and Astronomy, University of Southampton, Southampton, SO17 1BJ, United Kingdom}
\email{S.De-Liberato@soton.ac.uk}

\maketitle

\section*{ABSTRACT}
Much of the novel physics predicted to be observable in the ultrastrong light-matter coupling regime rests on the hybridisation between states with different numbers of excitations, leading to a population of virtual photons in the system's ground state. In this article, exploiting an exact diagonalization approach, we derive both analytical and numerical results for the population of virtual photons in presence of arbitrary losses. Specialising our results to the case of Lorentzian resonances we then show that the virtual photon population is only quantitatively affected by losses, even when those become the dominant energy scale. Our results demonstrate most of the ultrastrong-coupling phenomenology can be observed in loss-dominated systems which are not even in the standard strong coupling regime. We thus open the possibility to investigate ultrastrong-coupling physics to platforms that were previously considered unsuitable due to their large losses.

\section*{INTRODUCTION}
The study of the interaction between light and matter has been one of the cornerstones in the development of quantum mechanics. In most cases the light-matter coupling is weak enough to be intuitively described in terms of the emission and absorption of photons, while the matter system jumps between two of its quantised eigenstates. When the resonant coupling of an optically active transition with a mode of the electromagnetic field is larger than the losses determining their respective linewidths, it becomes possible to spectroscopically resolve the splitting due to the interaction. The system is then said to be in the strong light-matter coupling regime. Contrary to the weak coupling case, here the interaction between light and matter cannot be described in terms of emission and absorption of photons, but it is necessary to consider the dressed light-matter excitations of the coupled system. 
Finally if the coupling becomes even larger, comparable with the bare frequencies of the excitations, we enter a third regime, called ultrastrong coupling. Such a regime, described \cite{Ciuti05} and achieved \cite{Anappara09} for the first time using intersubband polaritons, has since been studied both theoretically and experimentally in a variety of different systems \cite{Baust15,Niemczyk10,Muravev11,Scalari12,Geiser12b,Schwartz11,Porer12,Askenazi14,Gubbin14,Gambino14,Maissen14,Zhang14,Goryachev14,Hagenmuller12,DeLiberato15}. Interest in this novel regime has been fuelled by its rich phenomenology,  including quantum phase transitions \cite{Lambert04,Ashhab13,Vukics14}, modification of energy transport \cite{Orgiu15,Feist15} and optical properties \cite{DeLiberato14,Ripoll15,Bamba15,Garziano17,Kockum17}, and the possibility to use it to influence chemical and thermodynamic processes \cite{Hutchison12,Hutchison13,Galego15,Cwik16,Cortese17}.

The relevant dimensionless parameter in a perturbative treatment of the light-matter interaction is the ratio between the coupling and the bare excitation frequencies. In the ultrastrong coupling regime such a parameter becomes non-negligible, with values larger than one recently achieved \cite{Yoshihara16}. Higher-order perturbative effects due to the antiresonant terms in the Hamiltonian, which do not conserve the number of excitations, are then able to hybridise states with different numbers of excitations. Such an hybridisation is at the origin of much of the ultrastrong-coupling phenomenology \cite{Bamba15,Garziano17,Kockum17} and one of its most striking consequences is that the ground state  $\ket{G}$ becomes a squeezed vacuum state containing a finite population of virtual photons. Those photons are said to be virtual because the ground state cannot radiate. Their presence can be however directly revealed when the system parameters are non-adiabatically modulated in time, transmuting virtual photons into real ones
\cite{Cirio16,Hagenmuller16,Stassi13,Garziano13,Dodonov06,DeLiberato07,DeLiberato09,Agnesi09,Faccio11,Carusotto12,Auer12,Fedortchenko16}, a process which presents strong analogies with the dynamical Casimir effect \cite{Lambrecht07,Johansson09,Nation12,Felicetti14,Stassi15} and with the Hawking radiation \cite{Finazzi13,Jacquet15,Steinhauser16}.
Non-adiabatic modulation of the parameters of a light-matter coupled system has been experimentally achieved in dielectric systems by modifying the dipole density with a femtosecond laser pulse \cite{Guenter09}, and in superconducting circuits  by applying an external flux bias \cite{Peropadre10,Wilson11}. Another promising proposal in this direction is the use of the superconducting to classical transition to alter the resonator response \cite{Scalari14}. 

While the best way to correctly model losses in the ultrastrong coupling regime has been object of much attention \cite{Beaudoin11,Bamba13,Bamba14b}, their impact on the structure of the ground state and on the presence of virtual photons has been for the moment almost totally neglected. One of the reasons is that interest in ultrastrong coupling has historically emerged from the study of strongly coupled systems. Its achievement is usually demonstrated by fitting the resonant splitting of the coupled resonances to measure the coupling strength.
Any system in which ultrastrong coupling has been demonstrated was thus {\it a fortiori} also in the strong coupling regime. But in this situation the loss rate is the smallest frequency scale of the problem, and perturbative methods that neglect its impact on the structure of the ground state are totally justified. Nevertheless strong and ultrastrong coupling depend on different figures of merit, and they are thus {\it a priori} independent regimes.
Systems in the ultrastrong but not in the strong coupling regime could still have large couplings, as well as large losses, both comparable with the bare frequency of the optical transition. The ground state of the system would then also hybridise
with its environment, modifying its structure. A thorough investigation of the effect of the losses on the virtual photon population in those systems then becomes necessary to ascertain whether ultrastrong-coupling phenomenology can still be observed or if  it is completely quenched by the effect of the environment.

In this article we calculate through a non-perturbative procedure the virtual photon population in presence of arbitrary losses. Specialising our results to the case of  Lorentzian light and matter resonances we prove that losses do have an impact on the virtual photon population, but only a quantitative one. Even in presence of dominant losses a sizeable fraction of virtual photons remains. Ultrastrong coupling phenomena can thus be observed in systems with very large couplings, in which losses have impeached the observation of strong coupling for intrinsic, or technological reasons. Prime examples could be graphene single and bilayers in which, notwithstanding different theoretical calculations predicting very large dipoles  \cite{Hagenmuller12,DeLiberato15}, strong coupling has not yet been achieved. Another example are hybrid quantum systems which were recently highlighted as ideal platforms for some quantum vacuum emission scheme \cite{Cirio16}. These have only very recently \cite{Mi16} achieved strong coupling, as they are characterised by large losses \cite{Toida13,Wallraff13}.
%%%%%%%%%%% FIG Disp %%%%%%%%%%%
\begin{figure}[t!]
%\vspace{-3.5cm}
\centering
\includegraphics[width=0.4\textwidth]{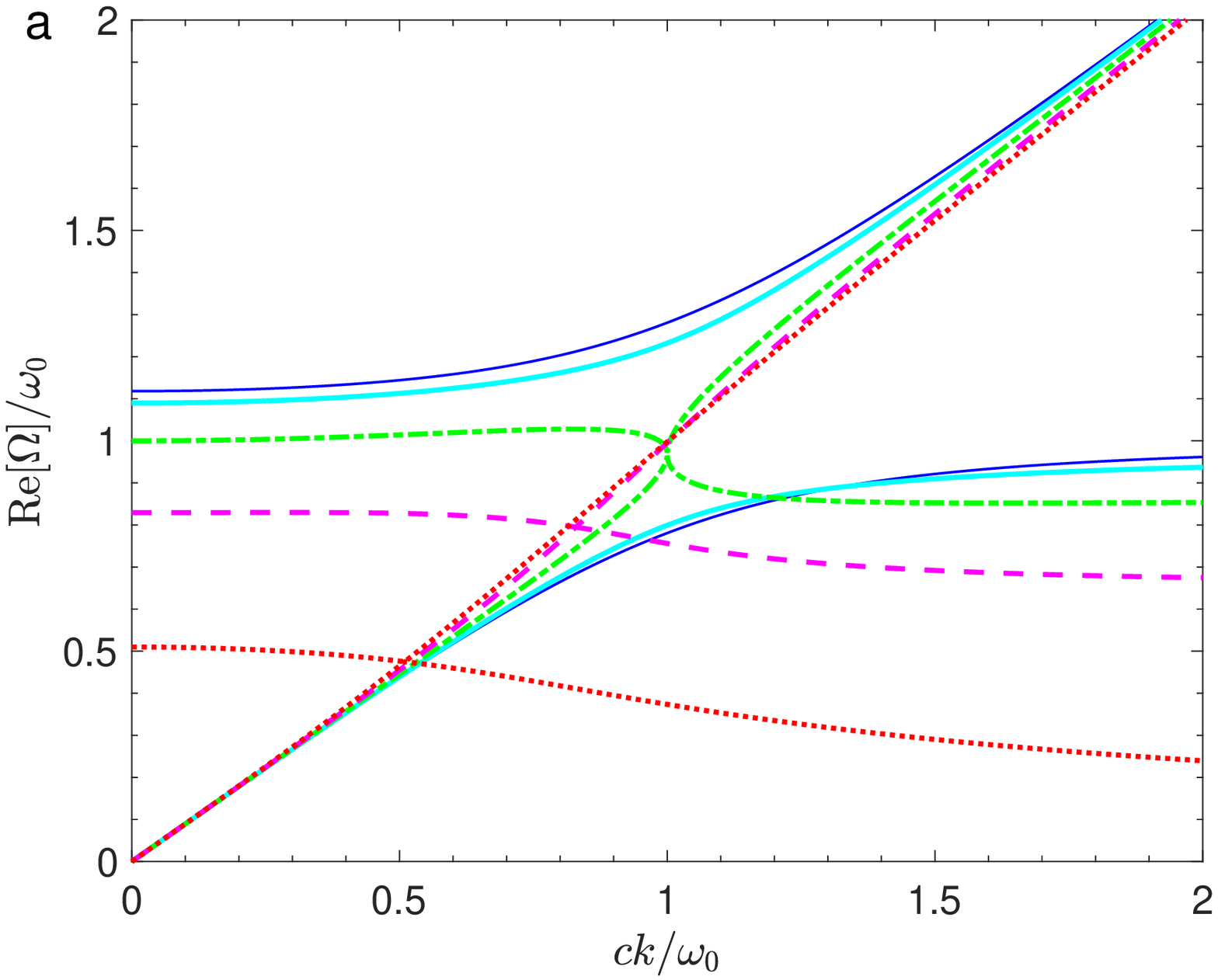}
\includegraphics[width=0.4\textwidth]{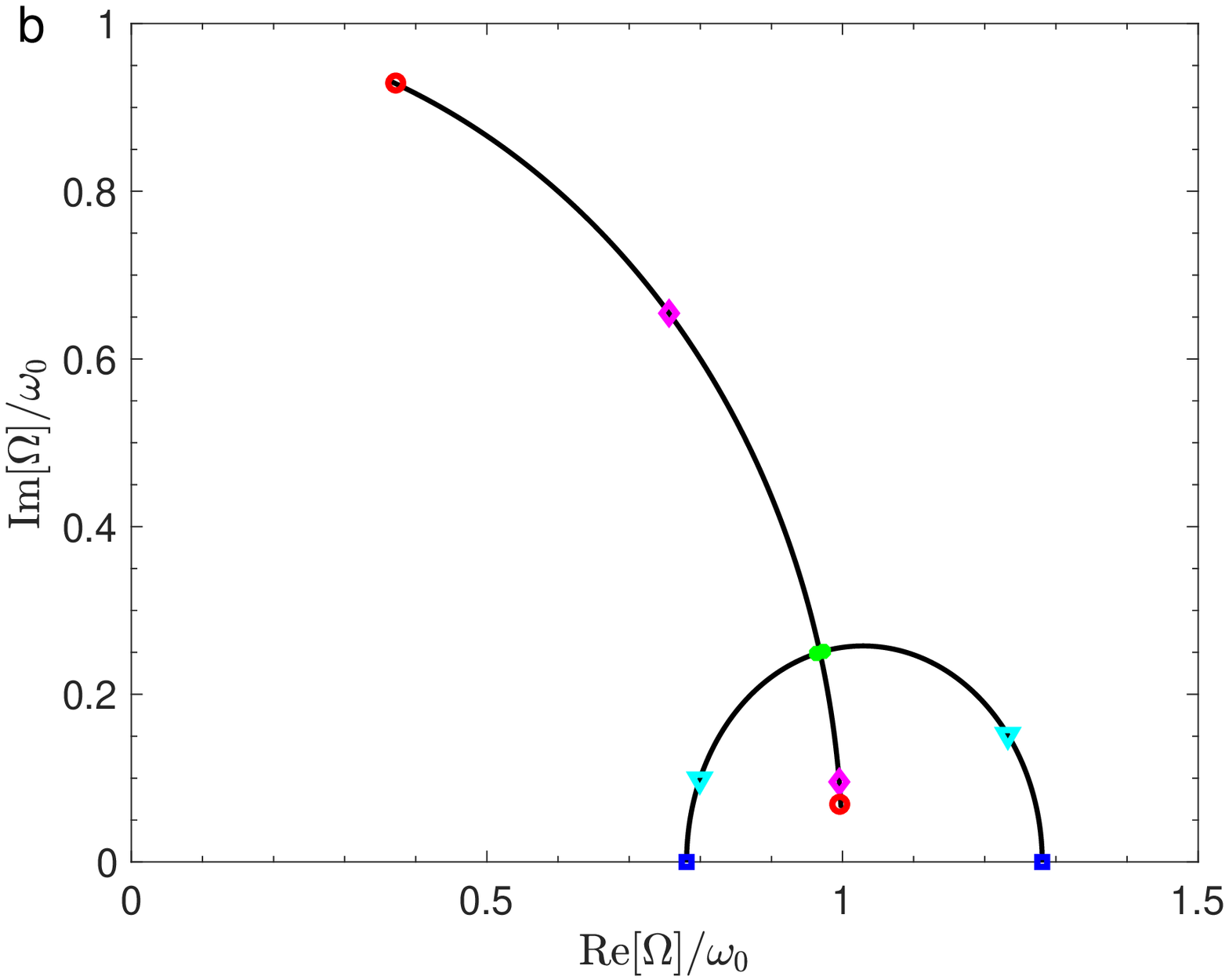}
  %\vspace{-3.5cm}
 \caption{\label{Fig1} Eigenmodes of the Lorentz model.
 {\bf a} Dispersion of the two polaritonic branches from the Lorentz model in \Eq{epsL}, for
 $\omega_{\text{c}}=0.5\omega_0$ and $\gamma_{\text{L}}=0$ (thin blue lines), $0.5\omega_0$ (solid cyan lines), $\omega_0$ (dash-dotted green lines), $1.5\omega_0$ (dashed magenta lines), and $2\omega_0$ (dotted red lines). The transition between the strong coupling regime presenting an anticrossing (blue and cyan lines) and the weak one in which the polaritonic modes cross (magenta and red lines), with the green line at the edge between the two, is cleary visible.
 {\bf b} Trajectories drawn by the eigenfrequencies in the complex plane, at resonance $ck=\omega_0$, when varying $\gamma_{\text{L}}$. Coloured 
 squares, triangles, dots, diamonds, and circles mark the increasing values of $\gamma_{\text{L}}$ used in {\bf a}.}
 \end{figure}
%%%%%%%%%%%%%%%%%%%%%%%%%

%%% Start theory %%%
\section*{RESULTS}
\section*{Analytical expression for the virtual photon population}
The quantity of interest for us will be the photonic population in the ground state $\ket{G}$ of the coupled light, matter, and environment fields. In our treatment this quantity is also the measure of the hybridisation between states with different numbers of excitations.
Due to the regime we are interested in, with all the parameters {\it a priori} of the same order, a perturbative approach would be unreliable and we are thus obliged to perform a non-perturbative calculation. 
In a light-matter coupled system energy can be lost through different channels. Photons can escape out of the system, or they can be absorbed by the matter excitation and their energy non-radiatively dissipated. The general theory we developed, detailed in the Supplementary Note 4, shows that in the considered parameter range the ground state photonic population essentially depends on the total amount of losses, regardless of their origin. Without loss of generality in the main body of the paper we will thus consider the case of losses due to absorption in a dielectric medium, which allows to obtain analytically intuitive results in terms of the complex dielectric function.   
In this case, considering an homogeneous dissipative dielectric with complex dielectric function $\epsilon(\omega)$, we show in the Methods section that the number of ground-state virtual photons in the mode $\mathbf{k}$ is given by
\begin{eqnarray}
\label{Nk}
N_{{k}}=
\sum_j
\text{Im}\left[\frac{{\Omega_j}^2-c^2k^2}{4\pi  c^3k^2}\frac{d{\Omega_j}}{dk}
\left(\I\pi-2\log({\Omega}_j)\right)\right]-\frac{1}{2},\;
\end{eqnarray}
where the $\Omega_j$ are the solution of the dispersion equation
\begin{eqnarray}
\label{dispersion}
\epsilon(\omega)\omega^2-c^2k^2&=&0,
\end{eqnarray}
located in the first quadrant of the complex plane.
As explained in Ref. \cite{Ciuti05}, $N_k$ is also the number of photons with wavevector $\mathbf{k}$ emitted upon an instantaneous switch-off of the interaction: after the switch-off the ground state would be the standard, empty vacuum, and all the virtual photons would be free to radiate. Notice that this identification remains valid in presence of losses because without light-matter coupling there can be no absorption and all the virtual photons in the ground are emitted. 

\section*{Numerical results for Lorentzian resonances}
In order to explore the physical content of \Eq{Nk} we apply it to a medium described by a dissipative Lorentz dielectric function
\begin{eqnarray}
\label{epsL}
\epsilon_{\text{L}}(\omega)=1+\frac{\omega_{\text{c}}^2}{\omega_0^2-\omega^2-\I\gamma_{\text{L}}\omega},
\end{eqnarray} 
which is a medium containing a single, dispersionless optically active resonance of frequency $\omega_0$, coupling strength $\omega_{\text{c}}$, and linewidth $\gamma_{\text{L}}$. 
It is well known that the spectrum of a medium described by \Eq{epsL} consists of two polaritonic branches, whose real frequencies cross or anticross accordingly to whether the system is in the weak ($\gamma_{\text{L}}>2\omega_{\text{c}}$) or in the strong  ($\gamma_{\text{L}}<2\omega_{\text{c}}$) coupling regime, as shown in Fig. \ref{Fig1}. 
With the appropriate choice of parameters, such a model can be used to describe, at least qualitatively, all linear dielectric condensed matter systems in which strong and ultrastrong coupling have been achieved to date. For historical reasons \cite{Anappara09} the threshold of ultrastrong coupling is usually assumed to be $\omega_{\text{c}}\geq0.2\omega_0$.
The poles of the dielectric function in \Eq{epsL}, 
$\Omega_0=\frac{-\I \gamma_{\text{L}}\pm\sqrt{4\omega_0^2-\gamma_{\text{L}}^2}}{2}$,
corresponding to the complex frequencies of the lossy matter resonance, have a real component only for $\gamma_{\text{L}}<\gamma_{\text{max}}=2\omega_0$. For $\gamma_{\text{L}}>\gamma_{\text{max}}$ the resonance thus becomes overdamped, the resonant frequency ill defined, and the analytic properties which allow to derive \Eq{Nk} don't apply anymore. Normally this is a sign that the dissipative Lorentz model is not adapted to describe the system under investigation and in the following we will thus take $\gamma_{\text{max}}$ as the largest physically meaningful value of the damping.

\begin{figure}[t]
%\vspace{-3.5cm}
\centering
\includegraphics[width=0.4\textwidth]{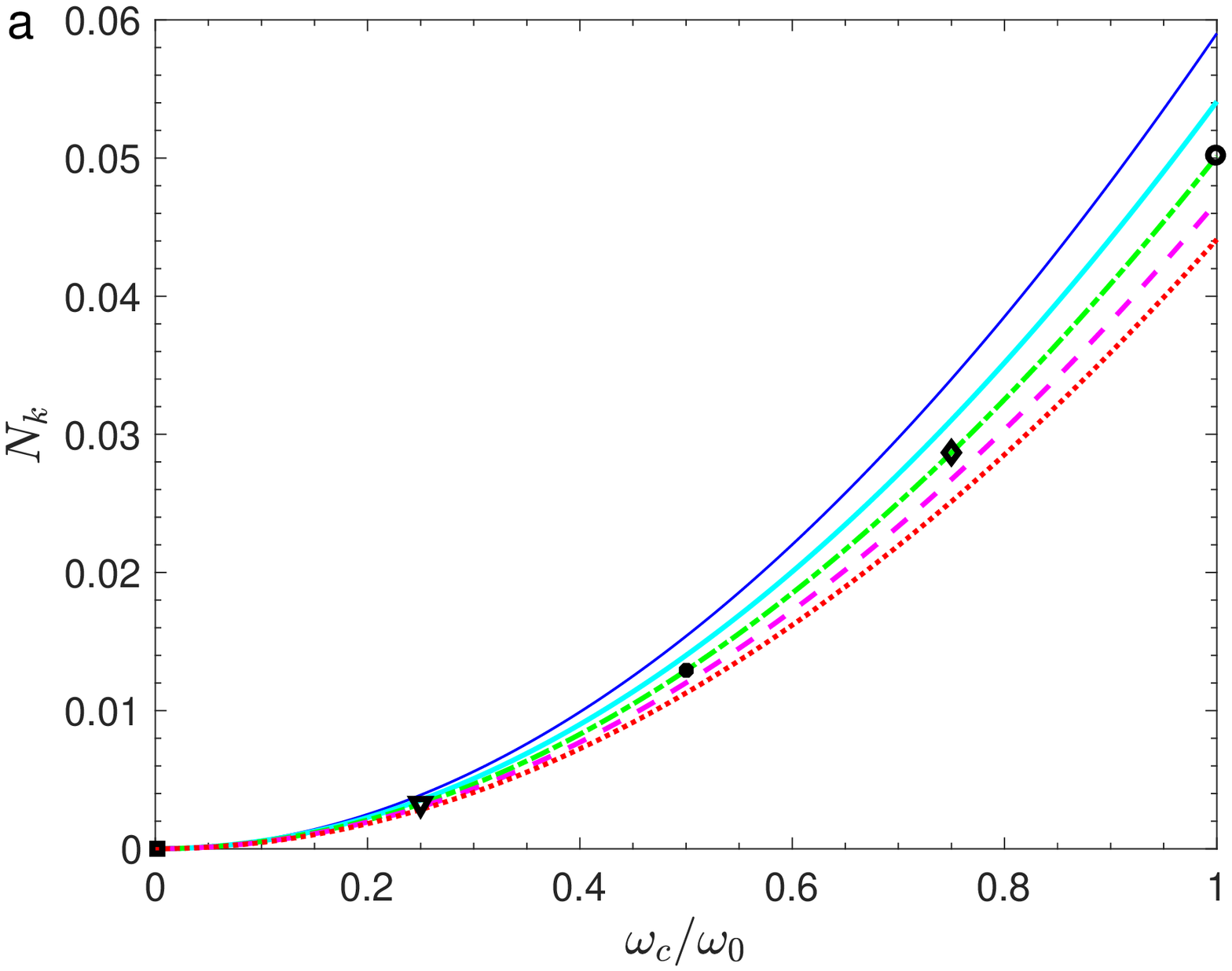}
\includegraphics[width=0.4\textwidth]{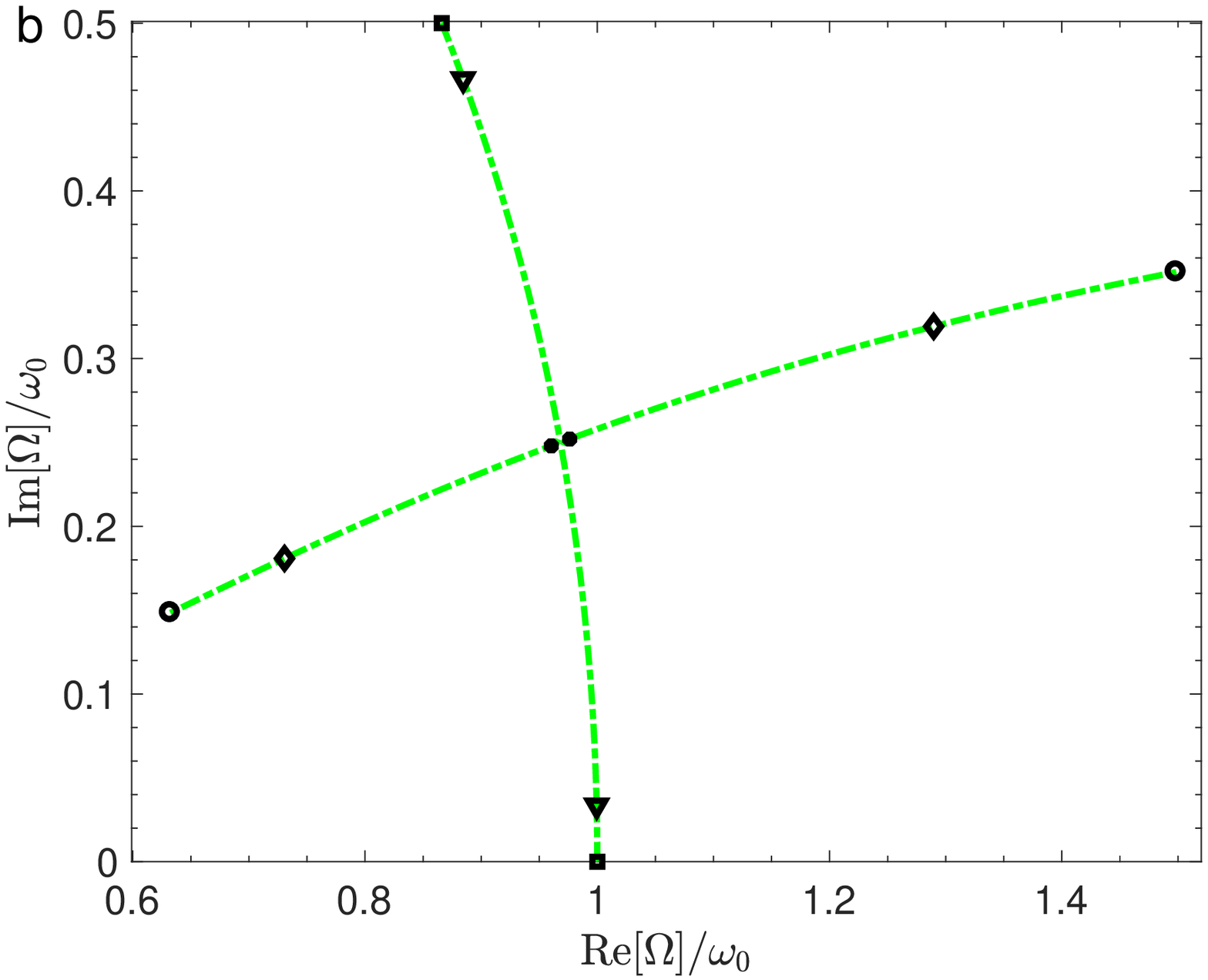}
  %\vspace{-3.5cm}
 \caption{\label{Fig2}
Virtual photons in the resonant Lorentz model. {\bf a}
Number of photons in the resonant mode $ck=\omega_0$ as a function of the coupling for $\gamma_{\text{L}}=0$ (thin blue line), $0.5\omega_0$ (solid cyan line), $\omega_0$ (dash-dotted green line), $1.5\omega_0$ (dashed magenta line), and $2\omega_0$ (dotted red line), that is the maximal physical value for the model we are considering. We see that going from the undamped to the overdamped regime, the number of virtual photons only diminishes of roughly the $25\%$. 
{\bf b}
Trajectory drawn by the eigenfrequencies in the complex plane, for $ck=\omega_0$ and $\gamma_{\text{L}}=\omega_0$, while varying $\omega_{\text{c}}$. The black symbols in {\bf a} and {\bf b} correspond to the same values of $\omega_{\text{c}}$. No visible discontinuity is present in the virtual photon population passing from the weak to the strong coupling regime.}
\end{figure}

In Fig. \ref{Fig2} we plot the number of virtual photons $N_k$ in the resonant mode $ck=\omega_0$ as a function of the light-matter coupling strength $\omega_{\text{c}}$. We recognize the expected, initially quadratic dependency over the coupling coefficient $\omega_{\text{c}}$ \cite{Ciuti05}.
Different lines relate to different values of $\gamma_{\text{L}}$ ranging from $0$ (thin blue line) to $\gamma_{\text{max}}$ (dotted red line). 
We see that, while losses do have a clear effect upon $N_k$, even in the case of an overdamped oscillator the virtual photon population is only diminished by roughly $25\%$ when compared with the nondissipative case.

The results in the case in which also a photonic linewidth $\gamma_{\text{P}}$ is present, obtained from the general approach developed in the Supplementary Note 4, can be found in the Supplementary Fig. 1. Those results show that also in this more general case our conclusions remain valid. A sizeable virtual photon population in fact remains, reduced at most by $50\%$ when $\gamma_{\text{L}}=\gamma_{\text{P}}=\gamma_{\text{max}}$ and the light and matter resonances are  both overdamped.
Moreover from Supplementary Fig. 1 we can see that in the considered parameter range $N_k$ essentially depends on the total linewidth $\gamma_{\text{L}}+\gamma_{\text{P}}$. The results in Fig. \ref{Fig2} thus generalise to this more general case if one considers the total linewidth instead that the matter one. 

As the coupling $\omega_{\text{c}}$ is varied from $0$ to $\omega_0$ we expect the system, at least for the small and intermediate values of $\gamma_{\text{L}}$, to transition from the weak to the strong coupling regime. Still no discontinuity is observed in $N_k$ showing that strong coupling has no direct effect on the virtual photon population. This can be confirmed from the inset of Fig. \ref{Fig2} where we plot the trajectory of the two complex polaritonic eigenfrequencies in the complex plane for $\gamma_{\text{L}}=\omega_0$, as $\omega_{\text{c}}$ is varied, identifying with different symbols specific values marked in the main image. We can clearly see a transition from the weak to the strong coupling regime as the two eigenfrequencies transition from having different loss rates but similar frequencies to the opposite case.
%%%%%%%%%%% FIG 1 %%%%%%%%%%%
%
%%%%%%%%%%%%%%%%%%%%%%%%%
%%%%%%%%%%% FIG 2 %%%%%%%%%%%
\begin{figure}[t]
%\vspace{-3.5cm}
\centering
\includegraphics[width=0.4\textwidth]{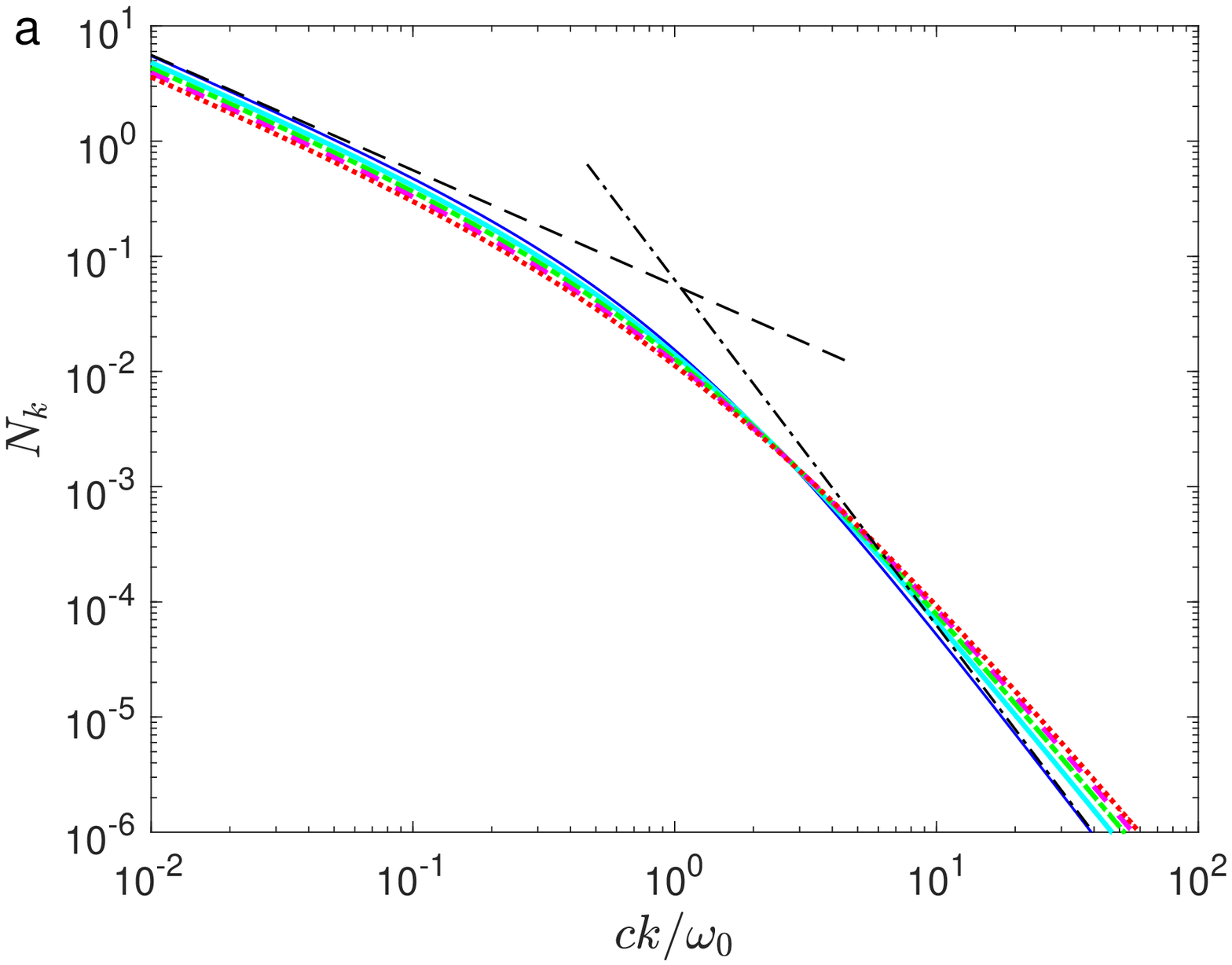}
\includegraphics[width=0.4\textwidth]{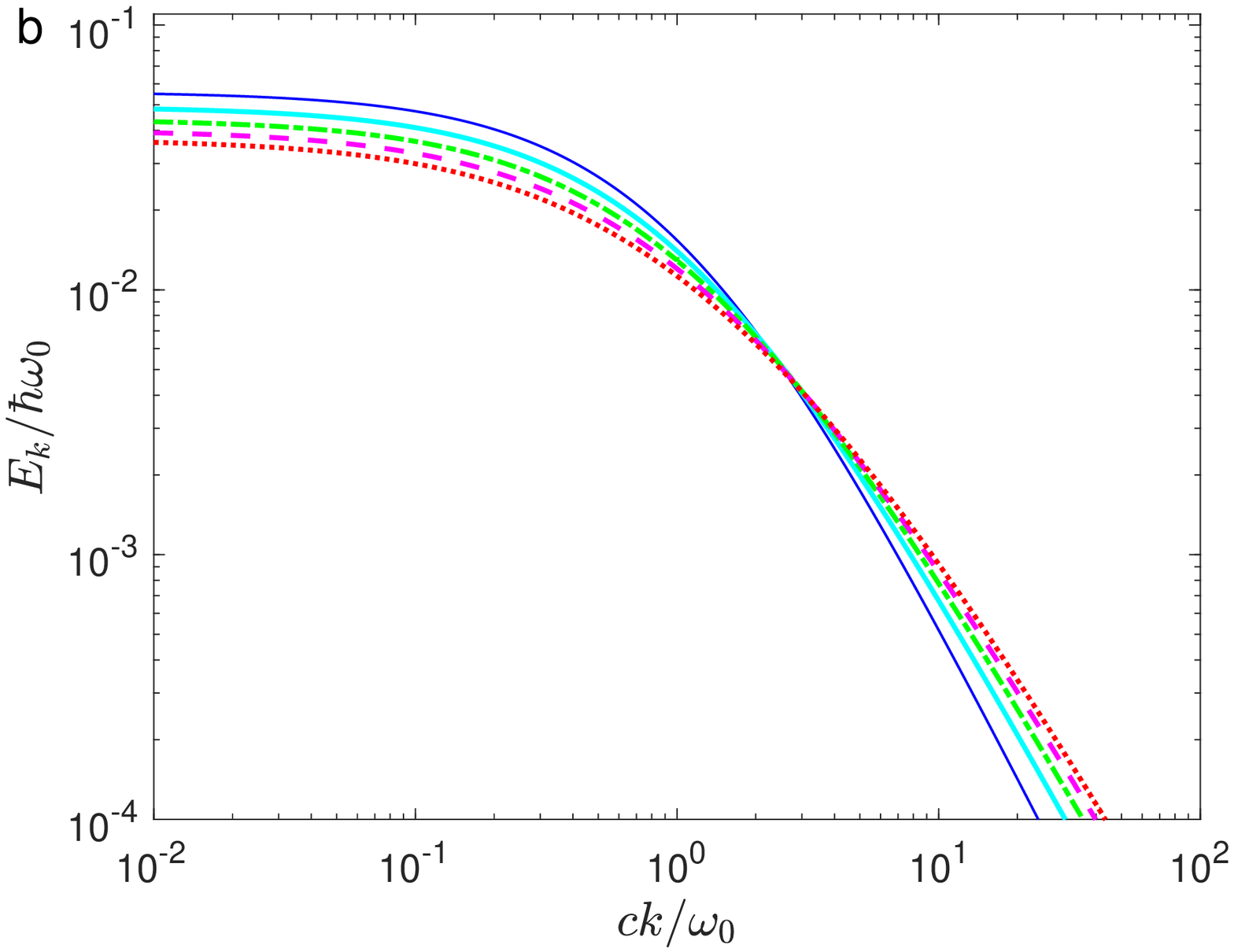} 
  %\vspace{-3.5cm}
 \caption{\label{Fig3} Virtual photons in the detuned Lorentz model.
{\bf a} Number of virtual photons in the ground state as a function of the photonic wavevector for $\omega_{\text{c}}=0.5\omega_0$  and different values of the losses. The black dashed and dash-dotted lines are the small and large $k$ expansions. {\bf b} Photonic energy per mode $E_k=\hbar ck N_k$ as a function of the photonic wavevector for $\omega_{\text{c}}=0.5\omega_0$  and different values of the losses. In both images $\gamma_{\text{L}}=0$ (thin blue line), $0.5\omega_0$ (solid cyan line), $\omega_0$ (dash-dotted green line), $1.5\omega_0$ (dashed magenta line), and $2\omega_0$ (dotted red line). }
 \end{figure}
%%%%%%%%%%%%%%%%%%%%%%%%%

\section*{Numerical results for strongly detuned systems}  
We verified that losses, even when larger than the light-matter coupling, have only a limited effect on the virtual population of the resonant photonic mode $ck=\omega_0$.
In order to ascertain if this also remains true out-of-resonance, in Fig. \ref{Fig3} we plot $N_k$ as a function of $\tfrac{ck}{\omega_0}$ over four orders of magnitude for a coupling $\omega_{\text{c}}=0.5\omega_0$ and values of the dissipation covering all the range between $0$ and $\gamma_{\text{max}}$. We verify again that dissipation does not have any qualitative impact, and also its quantitative effect is negligible for a plot over multiple orders of magnitude of the wavevector. More important, we do not see any sign of resonant enhancement of virtual photon population. This can be understood from the fact that the mixing of the vacuum state with states containing photonic excitations is due to the antiresonant terms of the Hamiltonian, and thus no resonance condition should be expected. Performing a perturbative development in inverse powers of $k$ (now justified as we are interested in extremal values of $k$) from of the dissipationless version of \Eq{Nk} we can find the asymptotic behaviours
\begin{eqnarray}
\label{Nklim}
N_{k\rightarrow 0}=\frac{\omega_{\text{c}}^2}{4ck\sqrt{\omega_0^2+\omega_{\text{c}}^2}},\quad\quad
N_{k\rightarrow \infty}=\frac{\omega_0\omega_{\text{c}}^2}{4c^3k^3},
\end{eqnarray}
plotted as black lines in Fig. \ref{Fig3}. Those results are consistent with the perturbative calculation in the dispersionless case performed in Ref. \cite{Beaudoin11}, predicting a larger squeezing for red-detuned resonators. They offer a first proof to a very recent conjecture by Roberto Merlin, linking the dynamical Casimir effect to the problem of orthogonality catastrophes, and predicting the presence of an infrared divergence in the number of generated low-energy photons \cite{Merlin16}.
In the inset of Fig. \ref{Fig3} we plot instead $E_k=\hbar ck\,N_k$, that is the photonic energy per mode, showing that it is also a monotonously decreasing function of the photonic wavevector $k$, saturating at 
$E_{\text{max}}=\frac{\hbar\omega_{\text{c}}^2}{4\sqrt{\omega_0^2+\omega_{\text{c}}^2}}$ for very red-detuned photons.\\

In summary, we demonstrated that the population of virtual photons present in the ground state of an ultrastrongly coupled system is solid against dissipation. Those results show that ultrastrong coupling physics can be observed in losses-dominated systems, in which strong coupling is not achievable.

\section*{METHODS}
\subsection*{Calculation of virtual photon population}
In order to derive the formula in \Eq{Nk} we can start by Huttner and Burnett's diagonalization method \cite{Huttner92}, which extended Hopfield's approach \cite{Hopfield58} to the case of a dispersive-dissipative dielectric. For sake of clarity we consider an homogeneous isotropic dielectric medium, although the extension to the inhomogeneous case does not present any fundamental issue \cite{Gubbin16}.  The derivation, detailed in the Supplementary Note 1, starts from an Hamiltonian describing the electromagnetic field coupled to an optically active transition. The latter is coupled to a reservoir of harmonic oscillators which act as a bath in which energy can be dissipated. Introducing annihilation operators $\hat{C}(\mathbf{k},\omega)$ for excitations of wavevector $\mathbf{k}$ and frequency $\omega$, obeying bosonic commutation relations 
\begin{eqnarray}
\left[ \hat{C}(\mathbf{k},\omega),
\hat{C}^{\dagger}(\mathbf{k'},\omega')
\right]=\delta(\mathbf{k-k'})\delta(\omega-\omega'),
\end{eqnarray}
and using a method originally due to Fano \cite{Fano56},  such an Hamiltonian can be put in the diagonal form 
\begin{eqnarray}
\label{Hdiag}
H=\sum_{\mathbf{k}} \int_0^{\infty} d\omega\; \hbar\omega\, \hat{C}^{\dagger}(\mathbf{k},\omega)\hat{C}(\mathbf{k},\omega).
\end{eqnarray}
The linear transformation used to diagonalize the system can then be inverted, allowing us to express the photonic operators as linear combinations of the $\hat{C}(\mathbf{k},\omega)$ as
\begin{eqnarray}
\label{aasC}
\hat{a}(\mathbf{k})=\int_0^{\infty} d\omega \left[ \tilde{\alpha}_{0,k}^*(\omega)\hat{C}(\mathbf{k},\omega)- \tilde{\beta}_{0,k}(\omega)\hat{C}^{\dagger}(-\mathbf{k},\omega)\right],
\end{eqnarray}
with
\begin{eqnarray}
\label{alphbeth}
\tilde{\alpha}_{0,k}(\omega)&=&\sqrt{\frac{\omega_{\text{c}}^2}{{ck}}}
\left(\frac{\omega+{ck}}{2}\right)
\frac{\zeta(\omega)}{\epsilon^*(\omega)\omega^2-c^2k^2},\quad\quad\quad\\
\tilde{\beta}_{0,k}(\omega)&=&\sqrt{\frac{\omega_{\text{c}}^2}{ck}}
\left(\frac{\omega-{ck}}{2}\right)
\frac{\zeta(\omega)}{\epsilon^*(\omega)\omega^2-c^2k^2},\nonumber
\end{eqnarray}
where the complex dielectric function is
\begin{eqnarray}
\label{eps}
\epsilon(\omega)&=&1+\frac{\omega_{\text{c}}^2}{2\omega}\int_{-\infty}^{\infty}d\omega'\,
\frac{\lvert\zeta(\omega')\rvert^2}{\omega'(\omega'-\omega-\I0^+)},
\end{eqnarray}
and the functional form of $\zeta(\omega)$ can be found in the Supplementary Note 1.
Exploiting the definition of ground state $\hat{C}(\mathbf{k},\omega)\ket{G}=0$, we can calculate the number of virtual photons with wavevector $\mathbf{k}$ as
\begin{eqnarray}
\label{Nkin}
N_{k}&=&\bra{G}\hat{a}^{\dagger}(\mathbf{k})\hat{a}(\mathbf{k})\ket{G}=\int_0^{\infty}d\omega\,\lvert \tilde{\beta}_{0,k}(\omega) \rvert^2
=\int_0^{\infty}d\omega\frac{(\omega-{k}c)^2}{2\pi {k}c}
\frac{\text{Im}\left[\epsilon(\omega)\right]\omega^2}{\lvert \epsilon(\omega)\omega^2-c^2k^2\rvert^2},
\end{eqnarray}
where $\text{Im}$ denotes the imaginary part, and we use \Eq{eps} and the Sokhotski-Plemelj theorem to write
\begin{eqnarray}
\text{Im}\left[\epsilon(\omega)\right]&=&\frac{\omega_{\text{c}}^2\pi\lvert\zeta(\omega)\rvert^2}{2\omega^2}.
\end{eqnarray}
Comparing \Eq{aasC} to \Eq{Nkin} we can verify that the ground state virtual photon population is the square of the mixing coefficient between annihilation and creation operators. The quantity in \Eq{Nkin} is thus a general measure of the hybridisation between states with different numbers of excitations, which is the key ingredient of most ultrastrong-coupling phenomenology \cite{Bamba15,Garziano17,Kockum17}.

As detailed below, the expression in \Eq{Nkin} can be calculated through an integral in the complex plane leading to the result in \Eq{Nk}. 
Note that the parameter $\omega_{\text{c}}$, which quantifies the light-matter coupling, can be  arbitrarily large. Our approach in fact takes into account the diamagnetic $\mathbf{A}^2$ term, forbidding the onset of superradiant phase transitions in polarizable media \cite{Rzazewski75,Bamba14}.
In order to verify our results, in the Supplementary Note 2 we compare the dissipationless limit of \Eq{Nk} to the formula obtained using the original Hopfield theory valid for nondissipative systems, showing that the two results coincide. Moreover in the Supplementary Note 3 we explicitly prove that \Eq{epsL} is in the form of the dielectric function in \Eq{eps}, and we can thus consistently apply it to \Eq{Nk}.

\subsection*{Calculation of the integral in the complex plane} 
\label{Integral} 
The total number of photons in the mode $\mathbf{k}$ from \Eq{Nkin}
can be calculated by noticing, following Ref. \cite{Huttner92}, that the dielectric function calculated at a complex frequency $\Omega$ satisfies the relation
\begin{eqnarray}
\label{epsilonproperty}
\epsilon(\Omega)&=&\epsilon^*(-\Omega^*),
\end{eqnarray}
and thus if $\Omega_j$ is a solution of \Eq{dispersion} so are $-\Omega_j,\Omega_j^*$, and $-\Omega_j^*$. 
Integrating over a keyhole contour in the complex plane, and developing the burdensome  algebra paying attention to chose the principal values of $\Omega_j$ to have the brach cut on the positive real axis, we arrive at
\begin{eqnarray}
N_{k}=\int_0^{\infty}d\omega\frac{(\omega-{ck})^2}{2\pi {ck}}
\frac{\text{Im}\left[\epsilon(\omega)\right]\omega^2}{\lvert \epsilon(\omega)\omega^2-c^2k^2\rvert^2}=\sum_j\left\{
\text{Im}\left[\frac{{\Omega_j}^2-c^2k^2}{4\pi  c^3k^2}\frac{d{\Omega_j}}{dk}
\left(\I\pi-2\log({\Omega}_j)\right)\right]-\text{Re}\left[\frac{\Omega_j}{2c^2k}\frac{d\Omega_j}{dk} \right]\right\},
\end{eqnarray}
where the sum is only over the solutions in the first quadrant.  Using the sum rule proved in Ref. \cite{Huttner92}
\begin{eqnarray}
\sum_j\text{Re}\left[\frac{\Omega_j}{k}\frac{d\Omega_j}{dk} \right]&=&c^2,
\end{eqnarray}
we then arrive to the final result in \Eq{Nk}.

\section*{ACKNOWLEDGEMENTS}
I wish to thank Krist\'in Arnard\'ottir, Motoaki Bamba, Neill Lambert, Ahsan Nazir, and Giacomo Scalari for useful discussions and feedback. I am Royal Society Research Fellow and I acknowledge support from EPSRC grant EP/M003183/1.

\newpage

%\maketitle
\renewcommand{\refname}{Supplementary References}
\def\bibsection{\section*{\refname}}

%%%%%%%%%%% FIG SUPP %%%%%%%%%%%
\begin{figure}[h!]
\centering
\includegraphics[width=0.49\textwidth]{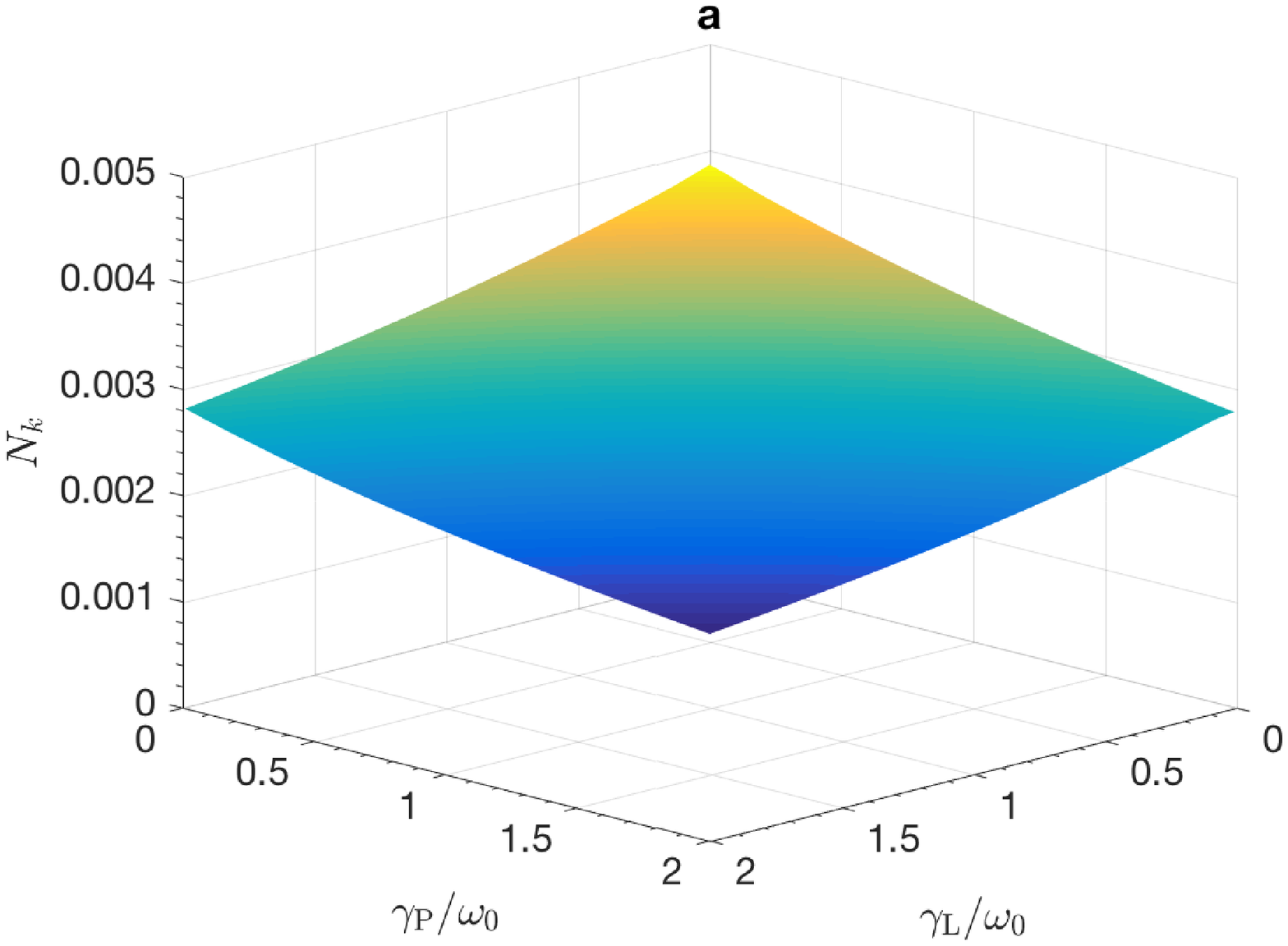}
\includegraphics[width=0.49\textwidth]{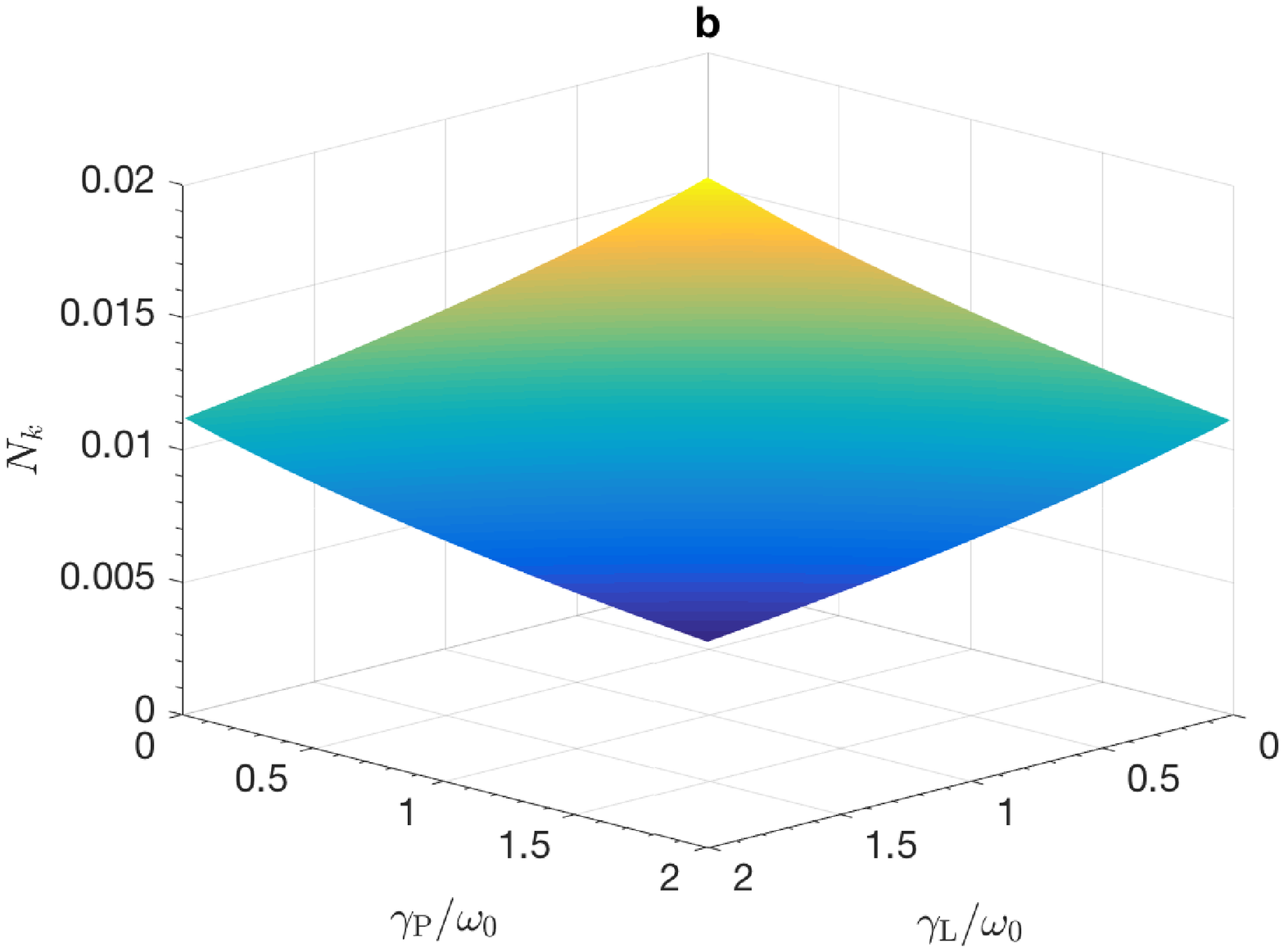}
\includegraphics[width=0.49\textwidth]{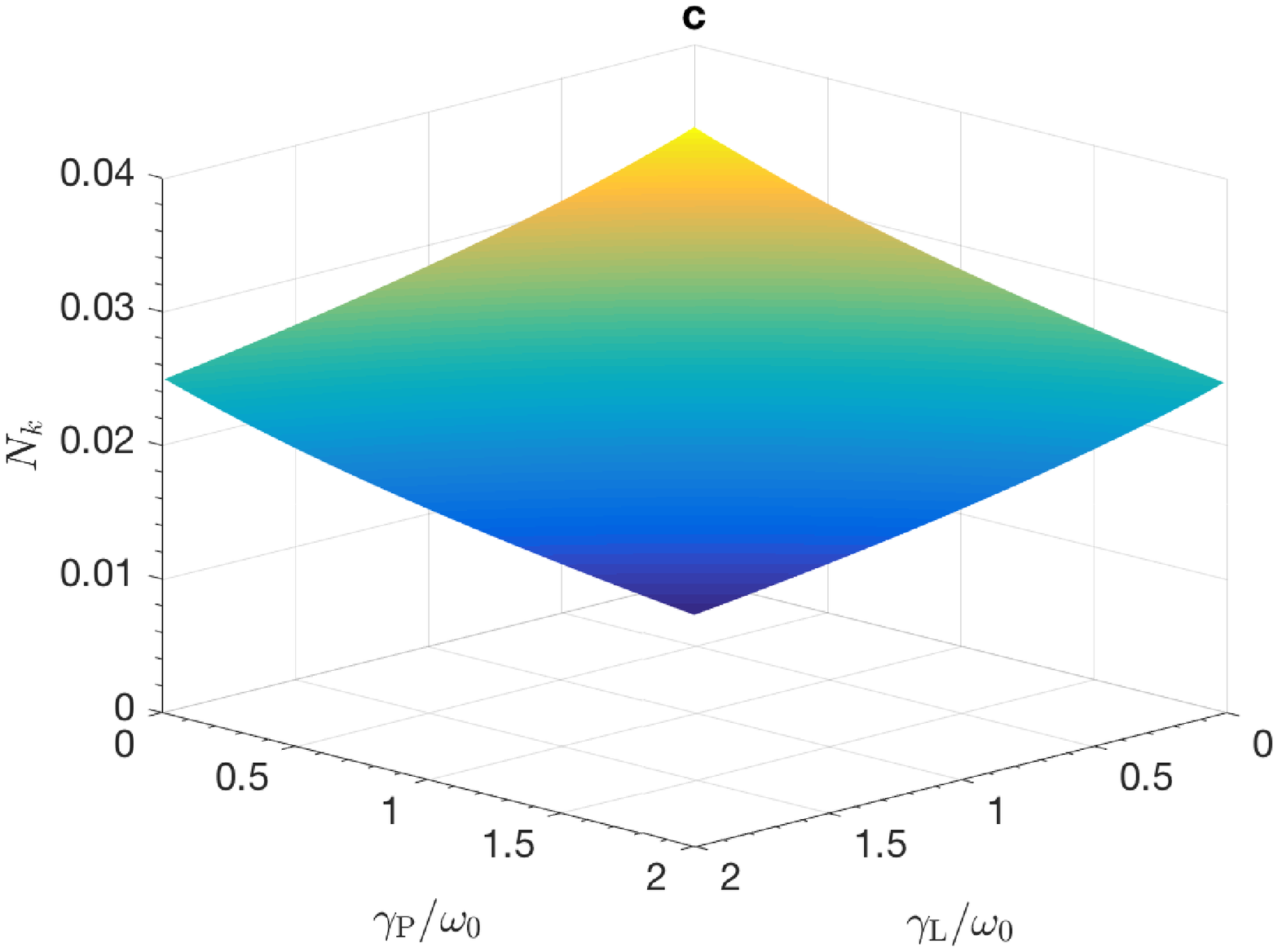}
\includegraphics[width=0.49\textwidth]{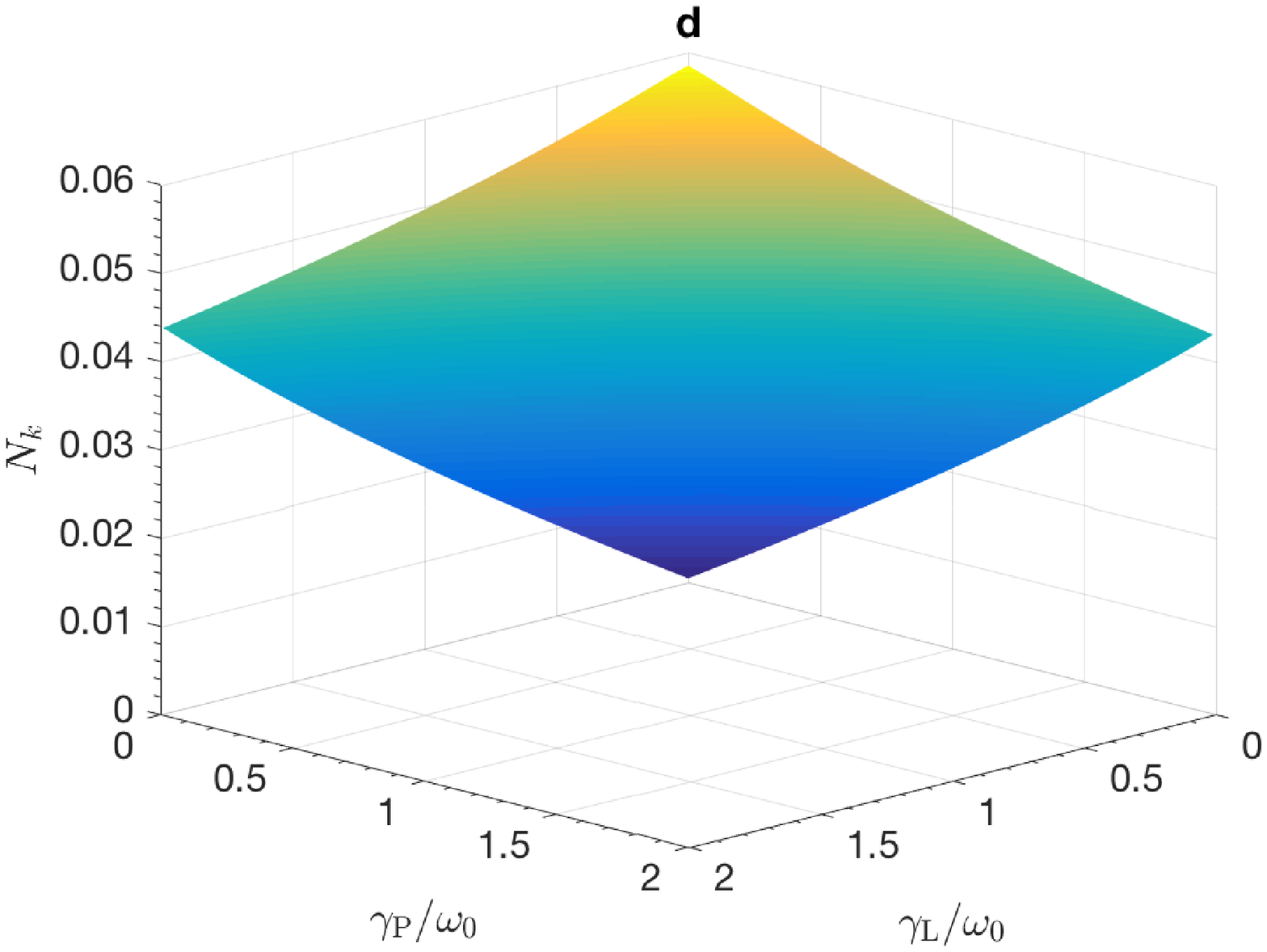}
 \end{figure}
%%%%%%%%%%%%%%%%%%%%%%%%%
\noindent
{\bf Supplementary Fig. 1:} Virtual photon population for arbitrary losses. Number of virtual photons in the resonant mode $ck=\omega_0$ as a function of the matter ($\gamma_{\text{L}}$) and photonic ($\gamma_{\text{P}}$) losses for different values of the light-matter coupling $\omega_{\text{c}}$. {\bf a} $\omega_{\text{c}}=0.25\omega_0$.
{\bf b} $\omega_{\text{c}}=0.5\omega_0$. {\bf c} $\omega_{\text{c}}=0.75\omega_0$. {\bf d} $\omega_{\text{c}}=\omega_0$.

\section*{Supplementary Note 1: Diagonalization of a dispersive-dissipative dielectric}
\label{HBprimer}
In their seminal work \cite{Huttner92} Huttner and Barnett extended the microscopic quantum theory of light-matter interaction developed by Hopfield \cite{Hopfield58} to the case of homogeneous but lossy dielectrics. They accomplished this by quantising the electromagnetic field coupled to a dispersionless matter excitation, itself coupled to a continuum reservoir of harmonic oscillators acting as a bath in which energy could be dissipated. 
The full Hamiltonian of such a system can be written as
\begin{eqnarray}
\label{HA}
\hat{H}&=&\hat{H}_{\text{em}}+\hat{H}_{\text{mat}}+\hat{H}_{\text{int}}+\hat{H}_{\mathbf{A}^2},
\end{eqnarray}
where
\begin{eqnarray}
\label{H1}
\hat{H}_{\text{em}}=\sum_{\mathbf{k}}\; \hbar {ck}\,\hat{a}^{\dagger}(\mathbf{k})\hat{a}(\mathbf{k}),
\end{eqnarray}
describes the free electromagnetic field,
\begin{eqnarray}
\label{H1M}
\hat{H}_{\text{mat}}=\sum_{\mathbf{k}}\; \left\{\hbar\tilde{\omega}_0\,\hat{b}^{\dagger}(\mathbf{k})\hat{b}(\mathbf{k})
\,+ \int_0^{\infty}d\omega\; \hbar\omega\, \hat{b}^{\dagger}_{\omega}(\mathbf{k})\hat{b}_{\omega}(\mathbf{k})
+ \int_0^{\infty}d\omega \;\frac{\hbar V(\omega)}{2}
  \left[\hat{b}^{\dagger}(-\mathbf{k})+\hat{b}(\mathbf{k}) \right]\left[\hat{b}^{\dagger}_{\omega}(\mathbf{k})+\hat{b}_{\omega}(-\mathbf{k}) \right]\right\},
\end{eqnarray}
models the matter excitation and the bath, 
\begin{eqnarray}
\hat{H}_{\text{int}}&=&\I\sum_{\mathbf{k}}\; \frac{\hbar\omega_{\text{c}}}{2}\sqrt{\frac{\tilde{\omega}_0}{ck}}\left[\hat{a}^{\dagger}(-\mathbf{k})+\hat{a}(\mathbf{k})\right]\left[\hat{b}^{\dagger}(\mathbf{k})-\hat{b}(-\mathbf{k}) \right],
\end{eqnarray}
is the dipolar interaction between light and matter, and
\begin{eqnarray}
\label{H4}
\hat{H}_{\mathbf{A}^2}&=&\sum_{\mathbf{k}}\; \frac{\hbar\omega_{\text{c}}^2}{4ck}  \left[\hat{a}^{\dagger}(-\mathbf{k})+\hat{a}(\mathbf{k})\right]\left[\hat{a}^{\dagger}(\mathbf{k})+\hat{a}(-\mathbf{k})\right],
\end{eqnarray}
comes from the diamagnetic $\mathbf{A}^2$ part of the the minimal-coupling Hamiltonian.
In Supplementary \Eqs{H1}{H4} $\hat{a}(\mathbf{k})$, $\hat{b}(\mathbf{k})$, and $\hat{b}_{\omega}(\mathbf{k})$, are bosonic annihilation operators respectively for a photon, a matter excitation, and an excitation of the bath with frequency $\omega$, all indexed by the wavevector $\mathbf{k}$, $\omega_0$ is the frequency of the optically active transition, $V(\omega)$ models its coupling to the bath, 
$\omega_{\text{c}}$ quantifies the intensity of the light-matter coupling, and the renormalised frequency $\tilde{\omega}_0$ is linked to the bare one by the formula
\begin{eqnarray}
\label{renormalisation}
\tilde{\omega}_0^2&=&\omega_0^2+\int_0^{\infty} \frac{\lvert V(\omega)\rvert^2\tilde{\omega}_0}{\omega}d\omega.
\end{eqnarray}
Note that we can chose arbitrary large values for the light-matter coupling $\omega_{\text{c}}$ because, thanks to the presence of the term $\hat{H}_{\mathbf{A}^2}$, the ground state is stable and no superradiant phase transition can take place, as demonstrated by various no-go theorems \cite{Rzazewski75,Bamba14}.

In Ref. \cite{Huttner92} the authors make the choice to remove $\hat{H}_{\mathbf{A}^2}$ by effectively performing a Bogoliubov rotation in the space of the $\hat{a}(\mathbf{k})$, thus putting $\hat{H}_{\text{em}}+\hat{H}_{\mathbf{A}^2}$ in diagonal form in terms of the Bogoliubov-rotated operators $\hat{\tilde{a}}(\mathbf{k})$ 
\begin{eqnarray}
\hat{\tilde{H}}_{\text{em}}=\sum_{\mathbf{k}}\; \hbar c\bar{k}\,\hat{\tilde{a}}^{\dagger}(\mathbf{k})\hat{\tilde{a}}(\mathbf{k}),
\end{eqnarray}
with $c\bar{k}=\sqrt{c^2k^2+\omega_{\text{c}}^2}$. Unluckily this is not acceptable for us, because it implies that the very definition of the the bare photonic operators depends upon the strength of the coupling, and as such it becomes problematic to define the number of virtual photons in the ground state. Ignoring this problem leads to a number of inconsistencies, most notably the total energy per mode diverges for $k\rightarrow 0$. 
In the following we will thus sketch a derivation that keeps $\hat{H}_{\mathbf{A}^2}$ as part of the interaction. 
The Hamiltonian in Supplementary \Eq{HA} can be diagonalised {\it a la} Fano \cite{Fano56} in two steps.
First $\hat{H}_{\text{mat}}$ is put into diagonal form 
\begin{eqnarray}
\label{DIAGMAT}
\hat{H}_{\text{mat}}=\sum_{\mathbf{k}} \int_0^{\infty} d\omega\; \hbar\omega\, \hat{B}^{\dagger}(\mathbf{k},\omega)\hat{B}(\mathbf{k},\omega),
\end{eqnarray}
where the $\hat{B}(\mathbf{k},\omega)$ operators, 
describing the continuously broadened optically active matter resonance, obey bosonic commutation relations
\begin{eqnarray}
\left[ \hat{B}(\mathbf{k},\omega),
\hat{B}^{\dagger}(\mathbf{k'},\omega')
\right]=\delta(\mathbf{k-k'})\delta(\omega-\omega').
\end{eqnarray}
They can be expressed as linear combinations of the different uncoupled matter operators as
\begin{eqnarray}
\hat{B}(\mathbf{k},\omega)=\alpha_0(\omega)\hat{b}(\mathbf{k})
+\beta_0(\omega)\hat{b}^{\dagger}(-\mathbf{k})
+\int_0^{\infty} d\omega'\;\left[\alpha_1(\omega,\omega')\hat{b}_{\omega'}(\mathbf{k})
+\beta_1(\omega,\omega')\hat{b}_{\omega'}^{\dagger}(-\mathbf{k})\right],
\end{eqnarray}
where the coefficients can be written as
\begin{eqnarray}
\alpha_0(\omega)&=&\left( \frac{\omega+\tilde{\omega}_0}{2}\right)\frac{V(\omega)}{\omega^2-\tilde{\omega}^2_0z(\omega)},\\
\beta_0(\omega)&=&\left( \frac{\omega-\tilde{\omega}_0}{2}\right)\frac{V(\omega)}{\omega^2-\tilde{\omega}^2_0z(\omega)},\\
\alpha_1(\omega,\omega')&=&\delta(\omega-\omega')+\left(\frac{\tilde{\omega}_0}{2}\right)
\left(\frac{V(\omega')}{\omega-\omega'-\I 0^+}\right)\frac{V(\omega)}{\omega^2-\tilde{\omega}^2_0z(\omega)},\\
\beta_1(\omega,\omega')&=&\left(\frac{\tilde{\omega}_0}{2}\right)
\left(\frac{V(\omega')}{\omega+\omega'}\right)\frac{V(\omega)}{\omega^2-\tilde{\omega}^2_0z(\omega)},
\end{eqnarray}
and 
\begin{eqnarray}
\label{z}
z(\omega)=1-\frac{1}{2\tilde{\omega}_0}\left[ \int_{-\infty}^{\infty}d\omega'  \frac{\lvert V(\omega')\rvert^2}{\omega'-\omega+\I 0^+} \right].
\end{eqnarray}
Introducing the quantity
\begin{eqnarray}
\zeta(\omega)=\I \sqrt{\tilde{\omega}_0}\left[\alpha_0(\omega)+\beta_0(\omega)\right]
=\I\frac{ \sqrt{\tilde{\omega}_0}\omega V(\omega)}{\omega^2-\tilde{\omega}^2_0z(\omega)},
\end{eqnarray}
which obeys the normalization
\begin{eqnarray}
\int_0^{\infty} d\omega \frac{\lvert \zeta(\omega)\rvert^2}{\omega}&=&1,
\end{eqnarray}
the interaction part of the Hamiltonian, now describing the interaction of the photonic modes with the broadened transition, takes the form
\begin{eqnarray}
\hat{H}_{\text{int}}&=&\sum_{\mathbf{k}} \;\frac{\hbar\omega_{\text{c}}}{2\sqrt{ck}}\int_0^{\infty}d\omega  \left\{ \left[\zeta(\omega)\hat{B}^{\dagger}(\mathbf{k},\omega)+\zeta^*(\omega)\hat{B}(-\mathbf{k},\omega)\right]\left[\hat{a}^{\dagger}(-\mathbf{k})+\hat{a}(\mathbf{k})\right]\right\}.
\end{eqnarray}
Analogously to the above, the novel Hamiltonian can now be put into diagonal form
\begin{eqnarray}
\hat{H}=\sum_{\mathbf{k}} \int_0^{\infty} d\omega\; \hbar\omega\; \hat{C}^{\dagger}(\mathbf{k},\omega)\hat{C}(\mathbf{k},\omega),
\end{eqnarray}
through the operators
\begin{eqnarray}
\hat{C}(\mathbf{k},\omega)=\tilde{\alpha}_{0,k}(\omega)\hat{a}(\mathbf{k})
+\tilde{\beta}_{0,k}(\omega)\hat{a}^{\dagger}(-\mathbf{k})
+\int_0^{\infty} d\omega'\;\left[\tilde{\alpha}_{1,k}(\omega,\omega')\hat{B}(\mathbf{k},\omega')
+\tilde{\beta}_{1,k}(\omega,\omega')\hat{B}^{\dagger}(-\mathbf{k},\omega')\right],
\end{eqnarray}
with the coefficients
\begin{eqnarray}
\label{tildecoeff1}
\tilde{\alpha}_{0,k}(\omega)&=&\sqrt{\frac{\omega_{\text{c}}^2}{ c{k}}}
\left(\frac{\omega+c{k}}{2}\right)
\frac{\zeta(\omega)}{\epsilon^*(\omega)\omega^2-c^2{k}^2},\quad\quad\quad\\
\label{tildecoeff2}
\tilde{\beta}_{0,k}(\omega)&=&\sqrt{\frac{\omega_{\text{c}}^2}{c{k}}}
\left(\frac{\omega-{ck}}{2}\right)
\frac{\zeta(\omega)}{\epsilon^*(\omega)\omega^2-c^2{k}^2},\\
\label{tildecoeff3}
\tilde{\alpha}_{1,k}(\omega,\omega')&=&\delta(\omega-\omega')
+\frac{\omega_{\text{c}}^2}{2}\frac{\zeta^*(\omega')}{\omega-\omega'-\I 0^+}
\frac{\zeta(\omega)}{\epsilon^*(\omega)\omega^2-c^2{k}^2},\\
\label{tildecoeff4}
\tilde{\beta}_{1,k}(\omega,\omega')&=&
\frac{\omega_{\text{c}}^2}{2}\frac{\zeta(\omega')}{\omega+\omega'}
\frac{\zeta(\omega)}{\epsilon^*(\omega)\omega^2-c^2{k}^2},
\end{eqnarray}
where the complex dielectric function is
\begin{eqnarray}
\label{epsHB}
\epsilon(\omega)=1+\frac{\omega_{\text{c}}^2}{2\omega}
\int_{-\infty}^{\infty}d\omega' \frac{\lvert \zeta(\omega') \rvert^2}{\omega'\left(\omega'-\omega-\I 0^+\right)}.
\end{eqnarray}
In comparison the equivalent of the coefficient in Supplementary \Eq{tildecoeff1}  obtained in Ref. \cite{Huttner92} has instead the form
\begin{eqnarray}
\tilde{\alpha}^{\text{HB}}_{0,k}(\omega)&=&\sqrt{\frac{\omega_{\text{c}}^2}{c\bar{k}}}
\left(\frac{\omega+c\bar{k}}{2}\right)
\frac{\zeta(\omega)}{\epsilon^*(\omega)\omega^2-c^2{k}^2},\quad\quad\quad
\end{eqnarray}
with the others following a similar scheme.

\section*{Supplementary Note 2: Virtual photon population in the lossless case}
\label{Hopfield}
In order to verify our calculations, it can be of use to compare the results obtained through the dissipative theory in the case of vanishing losses with those obtained using the standard nondissipative theory due to Hopfield \cite{Hopfield58}. In order to do this we start from the equivalent of Supplementary \Eq{HA} neglecting the bath 
\begin{eqnarray}
\label{HHop}
\hat{H}_{\text{Hop}}&=&\sum_{\mathbf{k}} \left\{ \hbar ck\, \hat{a}^{\dagger}(\mathbf{k})\hat{a}(\mathbf{k})+\hbar\omega_0\, \hat{b}^{\dagger}(\mathbf{k})\hat{b}(\mathbf{k})\right.\\&&\quad\left.
+\I\frac{\hbar\omega_{\text{c}}}{2}\sqrt{\frac{\omega_0}{ck}}\left[\hat{a}^{\dagger}(-\mathbf{k})+\hat{a}(\mathbf{k})\right]\left[\hat{b}^{\dagger}(\mathbf{k})-\hat{b}(-\mathbf{k})\right]
+\frac{\hbar\omega_{\text{c}}^2}{4ck}\left[\hat{a}^{\dagger}(-\mathbf{k})+\hat{a}(\mathbf{k})\right]
\left[\hat{a}^{\dagger}(\mathbf{k})+\hat{a}(-\mathbf{k})\right]\right\}
\nonumber.
\end{eqnarray}
The Hamiltonian in Supplementary \Eq{HHop} can be put in the diagonal form
\begin{eqnarray}
\hat{H}_{\text{Hop}}&=&\sum_{\mathbf{k}} \sum_{j=\pm} \hbar\omega_{j,k} \, \hat{p}^{\dagger}_{j}(\mathbf{k})\hat{p}_{j}(\mathbf{k}),
\end{eqnarray}
through the introduction of the polaritonic operators
\begin{eqnarray}
\label{pdef}
\hat{p}_{\pm}(\mathbf{k})&=&w_{\pm,k}\hat{a}(\mathbf{k})+x_{\pm,k}\hat{b}(\mathbf{k})+y_{\pm,k}\hat{a}^{\dagger}(-\mathbf{k})+z_{\pm,k}\hat{b}^{\dagger}(-\mathbf{k}),
\end{eqnarray}
where
\begin{eqnarray}
\label{evalue}
\omega_{\pm,k}=
\sqrt{\frac{\omega_0^2+\omega_{\text{c}}^2+c^2k^2\pm \sqrt{(\omega_0^2+\omega_{\text{c}}^2+c^2k^2)^2-4c^2k^2\omega_0^2} }{2}},
\end{eqnarray}
and 
\begin{eqnarray}
\label{evector}
\left(
\begin{array}{c}
  w_{\pm,k}   \\
  x_{\pm,k}  \\
  y_{\pm,k}   \\   
  z_{\pm,k}
\end{array}
\right)&=&\pm
\left\{ \frac{\omega_{\pm,k}}{\omega_0}\left[ \left(1- \frac{\omega^2_{\pm,k}}{\omega_0^2}\right)^2 +\frac{\omega_{\text{c}}^2}{\omega_0^2}\right]\right\}^{-\frac{1}{2}}
\left(
\begin{array}{c}
 \left[ \frac{\omega_{\pm,k}^2}{\omega_0^2}-1\right]\frac{\omega_{\pm,k}+ck}{2\omega_0}\sqrt{\frac{\omega_0}{ck}}  \\
  \I\frac{\omega_{\text{c}}}{2\omega_0}(1+\frac{\omega_{\pm,k}}{\omega_0}) \\
  \left[ \frac{\omega_{\pm,k}^2}{\omega_0^2}-1\right]\frac{\omega_{\pm,k}-ck}{2\omega_0}\sqrt{\frac{\omega_0}{ck}}  \\ 
   \I\frac{\omega_{\text{c}}}{2\omega_0}(1-\frac{\omega_{\pm,k}}{\omega_0}) 
\end{array}
\right).
\end{eqnarray}
The transformation from bare to polaritonic basis in Supplementary \Eq{pdef} can be inverted as
\begin{eqnarray}
\hat{a}(\mathbf{k})&=&w^*_{+,k}\hat{p}_+(\mathbf{k}) + w^*_{-,k}\hat{p}_-(\mathbf{k}) - y_{+,k}\hat{p}_+^{\dagger}(-\mathbf{k}) - y_{-,k}\hat{p}_-^{\dagger}(-\mathbf{k}),\\
\hat{b}(\mathbf{k})&=&x^*_{+,k}\hat{p}_+(\mathbf{k}) + x^*_{-,k}\hat{p}_-(\mathbf{k}) - z_{+,k}\hat{p}_+^{\dagger}(-\mathbf{k}) - z_{-,k}\hat{p}_-^{\dagger}(-\mathbf{k}),
\end{eqnarray}
and the photonic population in the mode $\mathbf{k}$ calculated 
with the Hopfield theory is thus
\begin{eqnarray}
\label{Np}
N'_{k}&=&\sum_{j=\pm} \lvert y_{j,k}\rvert^2.
\end{eqnarray}
We can now write $N_k$ in the limit of vanishing losses, and thus real eigenfrequencies, as
\begin{eqnarray}
\label{limN}
\lim_{V(\omega)\rightarrow 0} N_{{k}}=
\sum_{j=\pm}
\left[\frac{{\omega_{j,k}}^2-c^2k^2}{4  c^3k^2}\frac{d{\omega_{j,k}}}{dk}
\right]-\frac{1}{2}.
\end{eqnarray}
Plugging Supplementary \Eqs{evalue}{evector} into Supplementary \Eqs{Np}{limN}, and developing the heavy but straightforward algebra we can then prove that the two expressions coincide, verifying the correctness of our procedure.

\section*{Supplementary Note 3: Lorentz dielectric function}
\label{Lorentz}
Here we will prove that the Lorentz dielectric function 
\begin{eqnarray}
\label{epsL}
\epsilon_{\text{L}}(\omega)=1+\frac{\omega_{\text{c}}^2}{\omega_0^2-\omega^2-\I\gamma_{\text{L}}\omega},
\end{eqnarray} 
used through the paper is
consistent with the general form in Supplementary \Eq{epsHB}. We will do so by explicitly showing that the form in  Supplementary \Eq{epsL} is recovered by evaluating Supplementary \Eq{epsHB} with a coupling of the form
\begin{eqnarray}
\label{Vdef}
\lvert V(\omega)\rvert^2=\frac{\omega\tilde{\omega}_0}{q+\omega_M},
\end{eqnarray}
where $\omega_M$ is a cutoff frequency that we will eventually send to infinity and $q$ a positive constant frequency.
We will limit ourselves to consider the imaginary part of the dielectric function\begin{eqnarray}
\label{Imeps}
\text{Im}\left[\epsilon(\omega)\right]=\frac{\omega_{\text{c}}^2\,\tilde{\omega}_0^2\text{Im}\left[z(\omega)\right]}{(\omega^2-\tilde{\omega}_0^2\text{Re}\left[z(\omega)\right])^2+(\tilde{\omega}_0^2\text{Im}\left[z(\omega)\right])^2},
\end{eqnarray}
as the real part will be fixed by Kramers-Kronig relations.
Plugging Supplementary \Eq{Vdef} into Supplementary \Eq{renormalisation}, and introducing the cutoff in the integration we obtain
\begin{eqnarray}
\tilde{\omega}_0^2=\omega_0^2+\int_0^{\omega_M} \frac{\lvert V(\omega)\rvert^2\tilde{\omega}_0}{\omega}d\omega=\omega_0^2\frac{q+\omega_M}{q}.
\end{eqnarray}
From Supplementary \Eq{z}, using the Sokhotski-Plemelj theorem
\begin{eqnarray}
\text{Im}\left[z(\omega)\right]&=&\frac{\pi}{2}\frac{ \lvert V(\omega)\rvert^2}{\tilde{\omega}_0}=\frac{\pi}{2}\frac{\omega}{q+\omega_M},\\
\text{Re}\left[z(\omega)\right]&=&1-\frac{1}{2\tilde{\omega}_0}\mathcal{P}\int_{-\omega_M}^{\omega_M}d\omega' \frac{\lvert V(\omega')\rvert^2}{\omega'-\omega}\nonumber= 1-\frac{\omega_M}{q+\omega_M}+\frac{\omega\log\left(1-\frac{2\omega}{\omega+\omega_M}\right)}{2(q+\omega_M)},
\end{eqnarray}
where the $\mathcal{P}$ indicates the principal part of the integral.
Sending the cutoff to infinity we thus have 
\begin{eqnarray}
\label{limits1}
\lim_{\omega_M\rightarrow\infty}\tilde{\omega}_0^2\,\text{Im}\left[z(\omega)\right]&=&\frac{\pi}{2}\frac{\omega_0^2}{q}\omega, \\
\label{limits2}
\lim_{\omega_M\rightarrow\infty}\tilde{\omega}_0^2\,\text{Re}\left[z(\omega)\right]&=&\omega_0^2.
\end{eqnarray}
Using Supplementary \Eqs{limits1}{limits2} into Supplementary \Eq{Imeps} we finally get
\begin{eqnarray}
\text{Im}\left[\epsilon(\omega)\right]&=&\frac{\omega_{\text{c}}^2\,\omega_0^2\frac{\omega}{q}\frac{\pi}{2}}{(\omega^2-\omega_0^2)^2+(\omega_0^2\frac{\omega}{q}\frac{\pi}{2})^2},
\end{eqnarray}
which is the imaginary part of the Lorentz dielectric function
\begin{eqnarray}
\text{Im}\left[\epsilon_{\text{L}}(\omega)\right]=\frac{\omega_{\text{c}}^2\gamma_{\text{L}}\omega}{(\omega_0^2-\omega^2)^2+\gamma_{\text{L}}^2\omega^2},
\end{eqnarray}
upon the identification
\begin{eqnarray}
\label{gamma}
\gamma_{\text{L}}&=&\frac{\pi\omega_0^2}{2q}.
\end{eqnarray}
%%%
%\captionsetup{labelformat=empty}

%%%
\section*{Supplementary Note 4: Virtual photon population with arbitrary losses}
In the general situation in which photons can be lost both through absorption and leakage,
we need to consider a more general Hamiltonian 
\begin{eqnarray}
\label{HAP}
\hat{H}'&=&\hat{H}'_{\text{em}}+\hat{H}_{\text{mat}}+\hat{H}_{\text{int}}+\hat{H}_{\mathbf{A}^2},
\end{eqnarray}
with 
\begin{eqnarray}
\label{H1P}
\hat{H}'_{\text{em}}=\sum_{\mathbf{k}} \left\{ \hbar c\tilde{k}\,\hat{a}^{\dagger}(\mathbf{k})\hat{a}(\mathbf{k})
+ \int_0^{\infty}d\omega\; \hbar\omega\, \hat{a}^{\dagger}_{\omega}(\mathbf{k})\hat{a}_{\omega}(\mathbf{k})
+ \int_0^{\infty}d\omega \;\frac{\hbar Q_k(\omega) }{2}
   \left[\hat{a}^{\dagger}(\mathbf{k})+\hat{a}(-\mathbf{k})\right]
  \left[\hat{a}_{\omega}^{\dagger}(-\mathbf{k})+\hat{a}_{\omega}(\mathbf{k})\right]
  \right\}.
\end{eqnarray}
The Hamiltonian in Supplementary \Eq{H1P} is similar to the one in Supplementary \Eq{HA} but the external bath, modelling in this case a continuum of photonic modes represented by bosonic operators $\hat{a}_{\omega}(\mathbf{k})$, is now coupled through the coupling $Q_k(\omega)$ to the photonic excitation whose renormalised frequency is given by
\begin{eqnarray}
c^2\tilde{k}^2=c^2k^2+\int_0^{\infty} d\omega \frac{\lvert Q_k(\omega) \rvert^2 c\tilde{k}}{\omega}.
\end{eqnarray}
The Hamiltonian in Supplementary \Eq{H1P} can be put in the diagonal form
\begin{eqnarray}
\label{DIAGPHOT}
\hat{H}'_{\text{em}}=\sum_{\mathbf{k}} \int_0^{\infty} d\omega\; \hbar\omega\, \hat{A}^{\dagger}(\mathbf{k},\omega)\hat{A}(\mathbf{k},\omega),
\end{eqnarray}
in terms of the operators describing the continuum spectrum of a leaky resonator
\begin{eqnarray}
\hat{A}(\mathbf{k},\omega)=\check{\alpha}_{0,k}(\omega)\hat{a}(\mathbf{k})
+\check{\beta}_{0,k}(\omega)\hat{a}^{\dagger}(-\mathbf{k})
+\int_0^{\infty} d\omega'\;\left[\check{\alpha}_{1,k}(\omega,\omega')\hat{a}_{\omega'}(\mathbf{k})
+\check{\beta}_{1,k}(\omega,\omega')\hat{a}_{\omega'}^{\dagger}(-\mathbf{k})\right],
\end{eqnarray}
where the coefficients are
\begin{eqnarray}
\check{\alpha}_{0,k}(\omega)&=&\left( \frac{\omega+c\tilde{k}}{2}\right)\frac{Q_k(\omega)}{\omega^2-c^2\tilde{k}^2t(\omega)},\\
\check{\beta}_{0,k}(\omega)&=&\left( \frac{\omega-c\tilde{k}}{2}\right)\frac{Q_k(\omega)}{\omega^2-c^2\tilde{k}^2t(\omega)},\\
\check{\alpha}_{1,k}(\omega,\omega')&=&\delta(\omega-\omega')+\left(\frac{c\tilde{k}}{2}\right)
\left(\frac{Q_k(\omega')}{\omega-\omega'-\I 0^+}\right)\frac{Q_k(\omega)}{\omega^2-c^2 \tilde{k}^2t(\omega)},\\
\check{\beta}_{1,k}(\omega,\omega')&=&\left(\frac{c\tilde{k}}{2}\right)
\left(\frac{Q_k(\omega')}{\omega+\omega'}\right)\frac{Q_k(\omega)}{\omega^2-c^2\tilde{k}^2t(\omega)},
\end{eqnarray}
and 
\begin{eqnarray}
t(\omega)&=&1-\frac{1}{2c\tilde{k}}\left[ \int_{-\infty}^{\infty} d\omega' \frac{\lvert Q_k(\omega')\lvert^2}{\omega'-\omega+\I0^+}\right].
\end{eqnarray}
Introducing the quantity
\begin{eqnarray}
\chi_k(\omega)=\frac{1}{\sqrt{c\tilde{k}}}\left[\check{\alpha}_{0,k}(\omega)-\check{\beta}_{0,k}(\omega)\right]
=\sqrt{c\tilde{k}}\frac{Q_k(\omega)}{\omega^2-c^2\tilde{k}^2t(\omega)},
\end{eqnarray}
which obeys the normalization 
\begin{eqnarray}
\int_0^{\infty} d\omega\; \omega \lvert \chi_k(\omega) \rvert^2&=&1,
\end{eqnarray}
the interaction part of the Hamiltonian, describing two coupled continua, takes the form
\begin{eqnarray}
\hat{H}_{\text{int}}&=&\sum_{\mathbf{k}} \;\frac{\hbar\omega_{\text{c}}}{2}\int_0^{\infty}d\omega \int_0^{\infty}d\omega' 
\left[\chi_k(\omega)\hat{A}^{\dagger}(-\mathbf{k},\omega)+\chi_k^*(\omega)\hat{A}(\mathbf{k},\omega)\right]
\left[\zeta(\omega')\hat{B}^{\dagger}(\mathbf{k},\omega')+\zeta^*(\omega')\hat{B}(-\mathbf{k},\omega')\right],\\
\hat{H}_{\mathbf{A}^2}&=&\sum_{\mathbf{k}}\; \frac{\hbar\omega_{\text{c}}^2}{4}  \int_0^{\infty}d\omega \int_0^{\infty}d\omega'
\left[\chi_k(\omega)\hat{A}^{\dagger}(\mathbf{k},\omega)+\chi_k^*(\omega)\hat{A}(-\mathbf{k},\omega)\right]\left[\chi_k(\omega')\hat{A}^{\dagger}(-\mathbf{k},\omega')+\chi_k^*(\omega')\hat{A}(\mathbf{k},\omega')\right].
\end{eqnarray}
Contrary to the case treated previously, describing a single continuum coupled to a discrete resonance, the diagonalization of  two coupled continua was not treated in the original paper by Fano \cite{Fano56} and it does not seem to be readily available in the literature. We will thus describe the procedure in more details.
We aim to diagonalize the system in terms of two branches of continuum polaritonic modes
\begin{eqnarray}
\hat{P}_{\pm}(\mathbf{k},\omega)=\int_0^{\infty}d\omega' \left[ w_{\pm,k}(\omega,\omega')\hat{A}(\mathbf{k},\omega')+x_{\pm,k}(\omega,\omega')\hat{B}(\mathbf{k},\omega')+y_{\pm,k}(\omega,\omega')\hat{A}^{\dagger}(-\mathbf{k},\omega')+z_{\pm,k}(\omega,\omega')\hat{B}^{\dagger}(-\mathbf{k},\omega')\right]\hspace{-1mm},\;\;
\end{eqnarray}
obeying bosonic commutation relations
\begin{eqnarray}
\label{bosP}
\left[ \hat{P}_j(\mathbf{k},\omega),
\hat{P}^{\dagger}_{j'}(\mathbf{k'},\omega')
\right]=\delta(\mathbf{k-k'})\delta(\omega-\omega')\delta_{jj'}.
\end{eqnarray}
The eigenequation
\begin{eqnarray}
\omega \hat{P}_{\pm}(\mathbf{k},\omega)&=&\lbrack \hat{P}_{\pm}(\mathbf{k},\omega),\hat{H}'\rbrack,
\end{eqnarray}
leads to the system
\begin{eqnarray}
( \omega-\omega' )w_{j,k}(\omega,\omega')&=&
\omega_{\text{c}}^2\chi_k^*(\omega')\int d\omega'' \frac{\omega''}{\omega+\omega''}
 w_{j,k}(\omega,\omega'')\chi_k(\omega'')
+\omega_{\text{c}}\chi_k^*(\omega')\int d\omega'' \frac{\omega''}{\omega+\omega''} x_{j,k}(\omega,\omega'')\zeta(\omega''),\\
( \omega-\omega' )x_{j,k}(\omega,\omega')&=&
\omega_{\text{c}}\zeta^*(\omega')\int d\omega'' \frac{\omega''}{\omega+\omega''}
 w_{j,k}(\omega,\omega'')\chi_k(\omega''),
\end{eqnarray}
where the other two coefficients obey
\begin{align}
 & ( \omega-\omega' )w_{j,k}(\omega,\omega')\chi_k(\omega') =    ( \omega+\omega' )y_{j,k}(\omega,\omega')\chi_k^*(\omega'),
  \\  
  %&\nonumber\\
 & ( \omega-\omega' )x_{j,k}(\omega,\omega')\zeta(\omega') \;\;\,=      ( \omega+\omega' )z_{j,k}(\omega,\omega')\zeta^*(\omega').
\end{align}
Defining the functions
\begin{eqnarray}
\label{Kdef}
K_{j,k}(\omega)&=&\int d\omega' \frac{\omega' \chi_k(\omega')}{\omega+\omega'}w_{j,k}(\omega,\omega'),\\
W_k(\omega)&=& \mathcal{P}\int d\omega' \frac{\omega' \lvert \chi_k(\omega')\lvert^2}{\omega^2-\omega'^2},\\ 
Z(\omega)&=& \mathcal{P}\int d\omega' \frac{\omega^2\lvert \zeta(\omega')\lvert^2}{\omega'(\omega^2-\omega'^2)},
\end{eqnarray}
where $\mathcal{P}$ denotes the principal part, we can formally solve for the coefficients as
\begin{eqnarray}
\label{xdef}
x_{j,k}(\omega,\omega')\zeta(\omega')&=&\left[\mathcal{P}\frac{1}{\omega-\omega'}+s_{x,j,k}(\omega)\delta(\omega-\omega') \right]
\omega_{\text{c}}\lvert\zeta(\omega')\rvert^2K_{j,k}(\omega),\\
\label{wdef}
w_{j,k}(\omega,\omega')\chi_k(\omega')&=&\left[\mathcal{P}\frac{1}{\omega-\omega'}+s_{w,j,k}(\omega)\delta(\omega-\omega') \right]
\left[\frac{s_{x,j,k}(\omega)\lvert\zeta(\omega)\rvert^2}{2}
+Z(\omega)\right]\omega_{\text{c}}^2\lvert \chi_k(\omega')\lvert^2K_{j,k}(\omega),
\end{eqnarray}
with $s_{x,j,k}(\omega)$ and $s_{w,j,k}(\omega)$ functions to be determined.
Multiplying Supplementary \Eq{wdef} by $\frac{\omega'}{\omega+\omega'}$, integrating, and exploiting Supplementary \Eq{Kdef}, we finally get to the equation
\begin{eqnarray}
\label{careq}
1&=&\omega_{\text{c}}^2\left[\frac{s_{w,j,k}(\omega)\lvert \chi_k(\omega)\lvert^2}{2}+W_k(\omega) \right]
\left[
\frac{s_{x,j,k}(\omega)\lvert\zeta(\omega)\rvert^2}{2}+
Z(\omega)\right].
\end{eqnarray}
In order to fix the extra function introduced in Supplementary \Eqs{xdef}{wdef} we can now exploit Supplementary \Eq{bosP},
which after some algebra leads to 
\begin{eqnarray}
\label{normeq}
&& \bigg[ 
\omega_{\text{c}}^2\lvert\chi_k(\omega)\rvert^2\Big[\pi^2+  s^*_{w,j',k}(\omega)s_{w,j,k}(\omega) \Big]
\Big[
\frac{s_{x,j,k}(\omega)\lvert\zeta(\omega)\rvert^2}{2}
+Z(\omega)\Big]\Big[
\frac{s^*_{x,j',k}(\omega)\lvert\zeta(\omega)\rvert^2}{2}
+Z(\omega)\Big] \\&&
+\lvert\zeta(\omega)\rvert^2
\Big[\pi^2+  s^*_{x,j',k}(\omega)s_{x,j,k}(\omega) \Big]
\bigg]\nonumber  \omega_{\text{c}}^2 K_{j,k}(\omega) K^*_{j',k}(\omega)=\delta_{jj'}.
\end{eqnarray}
We have at this point determined five equations, two from Supplementary \Eq{careq} and three from Supplementary \Eq{normeq}, in the
six unknown functions $K_{j,k}(\omega), s_{w,j,k}(\omega), s_{x,j,k}(\omega)$, with $j=\pm$. 
The two eigenmodes $\hat{P}_{\pm}(\mathbf{k},\omega)$ are degenerate for each value of $\mathbf{k}$ and $\omega$.
This implies a basis in such a degenerate subspace needs to be chosen through gauge fixing, leading to a sixth equation which then allows to algebraically solve the system and complete the diagonalization.
We will fix this gauge freedom by choosing the basis in which $w_{+,k}(\omega,\omega)=0$,
which from Supplementary \Eq{wdef} implies $s_{w,+,k}=0$.
The number of ground state photons can at this point be readily determined by the formula
\begin{eqnarray}
\label{Nkfull}
N_k&=&\sum_{j=\pm} \int_0^{\infty}d\omega \int_0^{\infty}d\omega' \lvert y_{j,k}(\omega,\omega') \rvert^2
=\sum_{j=\pm} \int_0^{\infty}d\omega' \frac{ \omega_{\text{c}}^4 \lvert \chi_k(\omega')\lvert^2}{(\omega+\omega')^2}\int_0^{\infty}d\omega \;\lvert
\frac{s_{x,j,k}(\omega)\lvert\zeta(\omega)\rvert^2}{2}+
Z(\omega)\rvert^2 \lvert K_{j,k}(\omega)\lvert^2.\quad\quad
\end{eqnarray}

In order to get quantitative results from Supplementary \Eq{Nkfull} we now need to fix the coupling function $Q_k(\omega)$. The ultrastrong coupling regime has been achieved using many different kinds of photonic resonators, each one described by a different $Q_k(\omega)$. In order to achieve an acceptable level of generality, as done for the matter losses, we specialise the theory to the case of a Lorentzian resonance, which is usually an acceptable approximation for most real resonators in a quite broad parameter range.
We thus consider a coupling of the form of the one in Supplementary \Eq{Vdef}
\begin{eqnarray}
\label{Qdef}
\lvert Q_k(\omega)\rvert^2=\frac{\omega c\tilde{k}}{\tilde{q}+\omega_M},
\end{eqnarray}
with 
\begin{eqnarray}
\tilde{q}&=&\frac{\pi c^2k^2}{2\gamma_{\text{P}}}.
\end{eqnarray}
Calculating all the relevant integrals, and letting $\omega_M\rightarrow\infty$ at the end, we recover  
the normalised density for the photonic mode with broadening $\gamma_{\text{P}}$
\begin{eqnarray}
\omega \lvert \chi_k(\omega) \rvert^2
&=&\frac{2\gamma_{\text{P}} }{\pi}\frac{\omega^2}{(c^2k^2-\omega^2)^2+\gamma_{\text{P}}^2\omega^2}.
\end{eqnarray}
In the Supplementary Fig. 1 we plot the virtual photonic population of the resonant mode as a function of $\gamma_{\text{L}}$ and $\gamma_{\text{P}}$ for different values of $\omega_{\text{c}}$.
We can clearly see that in the considered parameter range $N_k$ essentially depends on the total broadening $\gamma_{\text{L}}+\gamma_{\text{P}}$, and that a sizeable virtual photon population, equal to roughly the $50\%$ of the lossless one, remains even when $\gamma_{\text{L}}=\gamma_{\text{P}}=\gamma_{\text{max}}$ and light and matter resonances are both overdamped.

\end{document}